The RSNA Lumbar Degenerative Imaging Spine Classification (LumbarDISC) Dataset

Authors:

Tyler J. Richards, Adam E. Flanders, Errol Colak, Luciano M. Prevedello, Robyn L. Ball, Felipe Kitamura, John Mongan, Maryam Vazirabad, Hui-Ming Lin, Anne Kendell, Thanat Kanthawang, Salita Angkurawaranon, Emre Altinmakas, Hakan Dogan, Paulo Eduardo de Aguiar Kuriki, Arjuna Somasundaram, Christopher Ruston, Deniz Bulja, Naida Spahović, Jennifer Sommer, Sirui Jiang, Eduardo Moreno Júdice de Mattos Farina,
Eduardo Caminha Nunes, Michael Brassil, Megan McNamara, Johanna Ortiz, Jacob Peoples, Vinson L Uytana, Anthony Kam, Venkata N.S. Dola, Daniel Murphy, David Vu, Dataset Contributor Group, Dataset Annotator Group, Competition Data Notebook Group, Jason F. Talbott

Affiliations
Department of Radiology, University of Utah School of Medicine (T.J.R., D.M.); Department of Radiology, Thomas Jefferson University, Philadelphia, PA (A.E.F.); Department of Medical Imaging, St. Michael's Hospital, Unity Health Toronto (E.C., H.M.L), Department of Radiology, The Ohio State University, Columbus, Ohio (L.M.P.), The Jackson Laboratory, Bar Harbor, ME (R.L.B.), Bunkerhilll Health, San Francisco, CA (F.K.); Department of Diagnostic Imaging, Universidade Federal de São Paulo (Unifesp), São Paulo, Brazil (F.K., E.M.J.d.M.F, E.C.N.); Department of Radiology and Biomedical Imaging, University of California San Francisco, San Francisco, CA (J.M., J.F.T); Informatics Department, Radiological Society of North America, Oak Brook, Ill (M.V.); University of Utah School of Medicine (A.K.); Radiology Department, Chiang Mai University Faculty of Medicine, Chiang Mai, TH (T.K., S.A.); Department of Radiology, Koc University School of Medicine, Istanbul, TR (E.A., H.D.); Department of Radiology, Icahn School of Medicine at Mount Sinai, New York, NY (E.A.); Diagnósticos da América S.A., São Paulo, BR (P.E.d.A.K); Gold Coast University Hospital, Gold Coast, Queensland, Australia (A.S., C.R); Clinical Center University of Sarajevo, Sarajevo, BA (D.B., N.S); Diagnostic Imaging and Radiology, University Hospitals Cleveland Medical Center, Cleveland, OH (J.S., S.J.); Tallaght University Hospital Radiology Department, Dublin, IE (M.B., M.M); Kingston Health Sciences Centre, Kingston, ON, CA (J.O.); School of Computing, Queen's University, Kingston, ON, CA (J.P); Department of Imaging, Cedars-Sinai Medical Center, Los Angeles, CA (V.L.U.); Departments of Radiology and Neurosurgery, Loyola University Medical Center, Chicago, IL (A.K); George Washington University Hospital, Washington, DC (V.N.S.D); Children's National Hospital, Washington, DC (V.N.S.D); Scripps Clinical Medical Group,  San Diego, CA (D.V.)

Dataset Contributor Group

Samantha Yost - University of Utah School of Medicine
Nedim Kruscica - Berg d.o.o Sarajevo
Damir Hadzalic MD - Clinical center University of Sarajevo
Rahin Chowdhury M.S. - University Hospitals Cleveland Medical Center
Conor Waters MB BAO BCh, FFR RCSI, PGDip - Tallaght University Hospital
Dr Johnathon Harris BM MCh BSc, MRCSI - Tallaght University Hospital
Arsalan P. Rizwan MD, PhD - Queen's University Health Sciences, Kingston Health Sciences Centre
Jianwei Yue M.Sc. - Queen's University

Dataset Annotator Group

Arsany Hakim, MD - Bern University Hospital, Inselspital
Lai Peng Chan, FRCR, MBBS - Singapore General Hospital
Vinson Louis Uytana, MD - Cedars-Sinai Medical Center
Anthony Kam, MD, PhD - Loyola University Medical Center
Venkata Naga Srinivas Dola, DM, FRCR - Children's National Hospital, George Washington University
Girish Bathla, MD, FRCR - Associate Professor, Mayo Clinic, Rochester
Yonghan Ting, FRCR - National University Hospital, Singapore
Daniel Murphy, MD - University of Utah
David Vu, MD - Scripps Clinic Medical Group
Gagandeep Choudhary, MD, MBBS - Oregon Health and Science University
Tze Chwan Lim, FRCR, MBBS - Woodlands Health
Luciano Farage, MD - UNIEURO
Christie Lincoln, MD - MD Anderson Cancer Center
Kian Ming Chew, MBChB - Woodlands Health Singapore
Katie Bailey, MD - University of South Florida
Eduardo Portela de Oliveira, MD - The Ottawa Hospital, University of Ottawa
Fanny Moron, MD - Baylor College of Medicine
Achint Kumar Singh , MD - UT Health San Antonio
Nico Sollmann, MD, PhD - University Hospital Ulm
Kim Seifert, MD, MS - Stanford
Eric D. Schwartz, MD - Director of Neuroradiology, St. Elizabeth's Medical Center
Mariana Sanchez Montaño, MD - Rh Radiologos
Charlotte Yuk-Yan Chung, MD, PhD - NYU Langone Health
Lubdha Shah , MD - University of Utah

Ling Ling Chan, FRCR, MBBS - Singapore General Hospital
Scott R. Andersen, MD - Colorado Kaiser
Troy Hutchins, MD - University of Utah
Rita Nassanga, Mmed Radiology, MBChB - Makerere University, Kampala Uganda
Rukya Ali Masum - Ohio State Wexner Medical Center
Karl Soderlund, MD - Naval Medical Center Portsmouth
Le Roy Chong, MBBS, FRCR - Changi General Hospital
Jonathan D. Clemente, MD - Carolinas Medical Center
Ali Haikal Hussain, FRCR, MBChB - University of Rochester
Keynes Low - Woodlands Health
Mohiuddin Hadi, MD - University of Louisville
Michael Hollander, MD - Danbury Radiology Associates
Nurul Hafidzah Binti Rahim, MD - Hospital Putrajaya, Malaysia
Angela Guarnizo Capera, MD - Fundación Santa Fe de Bogotá
Lex A. Mitchell, MD - Hawaii Permanente Medical Group
Gennaro D'Anna, MD - ASST Ovest Milanese
Ellen Hoeffner, MD - University of Michigan
John L. Go, MD - University of Southern Califotnia
Facundo Nahuel Diaz, MD - Atrys Health / Hospital Italiano de Buenos Aires
Jacob Ormsby, MD, MBA - University of New Mexico
Jaya Nath, MD - Northport VA Medical center
Nathaniel von Fischer, MD - Kaiser Permanente South San Francisco
Vahe M. Zohrabian, MD - Northwell Health, North Shore University Hospital
Mary Niroshinee Muthukumarasamy, MBBS, MD - Ministry of Health, Sri Lanka
Sucari Vlok, MBChB, MMed - Tygerberg Hospital, University of Stellenbosch
Nafisa Paruk, FCRad diagnostics, 2SA, MBChB – Dr. Oosthuizen and Partners
Shayan Sirat Maheen Anwar, MBBS, FCPS - Aga Khan University Hospital
Giuseppe Cruciata, MD - Stony Brook University Hospital
Omar Islam, MD, FRCPC - Queen's University
Loizos Siakallis , MD - University College London
Ichiro Ikuta, MD, MMSc - Mayo Clinic Arizona

Competition Data Notebook Group

Abhinav Suri MPH - University of California Los Angeles
Andrew Wentland MD PhD - University of Wisconsin
Hari Trivedi MD - Emory University

Introduction

Low back pain is a worldwide health problem with lifetime prevalence reported as high as 84%.[1] In the United States alone, the direct medical and indirect costs of low back pain have been estimated at $50-100 billion.[2] In the lumbar spine, narrowing of the spinal canal, neural foramina, and subarticular recesses can compress spinal nerve roots, which can lead to radiculopathy and necessitate spinal surgery. MRI is the best imaging modality to assess the degree of narrowing in these regions, but reported inter-rater agreement between radiologists' grades of stenosis has been variable, ranging from poor to substantial, between different studies and spinal location (spinal canal, subarticular, or neural foramen).[3,4] While the large number of lumbar spine MRIs creates a substantial workload for radiologists, this is overshadowed by the more critical challenge of ensuring consistent and accurate grading of stenosis severity, which impacts diagnostic confidence and potentially patient care. Previous efforts to create machine learning (ML) models for this purpose have been limited by the available lumbar spondylosis imaging datasets.[5–7] Our goal was to create the largest, most diverse, expertly annotated MRI lumbar spondylosis dataset for both the RSNA 2024 Lumbar Spine Degenerative Classification AI Challenge and future research.

Dataset Description and Usage

The RSNA Lumbar Degenerative Imaging Spine Classification (LumbarDISC) dataset is composed of MRI studies of the lumbar spine from 2,697 patients with a total of 8,593 image series from 8 institutions across 6 countries and 5 continents. Inclusion criteria included an MR lumbar spine study with a sagittal "T2-like" (e.g. conventional spin echo T2, STIR, or Dixon), sagittal T1, and axial T2 weighted images. Both axial T2 images acquired as a continuous stack or oriented parallel to the intervertebral disc levels were acceptable if they included at least 3 lumbar spine disc levels. Additional inclusion criteria were a patient age ≥ 18 years old and imaged in an outpatient setting for lumbar degenerative disease. Exclusion criteria included lumbosacral spinal hardware, active non-degenerative pathology (e.g. spinal tumor, active infection, etc.), severe scoliosis, or substantial diagnostic limiting artifact.

A summary of the patient demographics and incidence of high-grade (moderate or severe) degenerative stenosis by institution is included in Table 1. Overall, of the 13,474 spinal canal stenosis grades, 85.4% were normal/mild, 8.8% were moderate, and 5.9% were severe. Of the 26,919 neural foraminal grades, 78.3% right and 77.2% left were normal/mild, 17.3% right and 18.1% left were moderate, and 4.4% right and 4.6% left were severe. Of the 26,285 subarticular grades, 69.4% right and 69.5% left were normal/mild, 19.6% right and 19.2% left were moderate, and 10.9% right and 11.3% left were severe. Distribution of high-grade degenerative stenosis across contributing institutions and dataset partitions (training, private, and public test sets) are provided in Tables 1 and 2.

Figure 1 and Appendix A describe the annotation and curation process as a flowchart and in more detail, respectively. Briefly, imaging and demographic data were acquired from each contributing site as detailed in Appendix B. Volunteer annotators from the Radiological Society of North America (RSNA), American Society of Neuroradiology (ASNR), and the American Society of Spine Radiology (ASSR) graded the degree of stenosis (4 point scale of normal, mild, moderate, or severe) in the following 5 locations: spinal canal (labeled on the T2-like sequence), right and left neural foraminal (sagittal T1), and right and left subarticular recess (axial T2) (Figure 2). A localizer was also placed for each annotation, centered at the spinal level corresponding to the assessed stenosis grade (e.g. centered within the L1/L2 neural foramen for the stenosis grade at that level) (see Figure 1). Annotators were provided an instruction manual (Appendix C) and a short instructional video to review. Each annotator was assigned to annotate one location (spinal canal, neural foramina, or subarticular stenosis) based on their performance on each location on a 10 case practice test. After all cases were annotated once, the 4-point severity scale was collapsed into a 3-point scale by combining normal and mild. From this, studies were divided into training (n=1,981), public test (n=272), and private test (n=444) data sets. Emphasis was placed on optimally distributing the data with respect to sex, age, contributing site, and disease severity classes across the training, public test, and private test sets. A greater portion of the high-grade stenosis cases at L1/L2, L2/L3, and L5/S1 were distributed to the test sets to have adequate examples to evaluate the models at these naturally under-represented levels. We also had 3 data contributing sites (sites 2, 4, and 7) search their database to identify additional cases of high grade disease at these levels. For the test datasets, an additional 1 to 3 annotations were acquired until 2 annotators agreed on a degree of severity which established the consensus grade. See Figure 3 for severity distributions of high grade stenosis by training and test sets and Appendix A for more detailed description of the annotation process.

The final dataset is available at [https://www.kaggle.com/competitions/rsna-2024-lumbar-spine-degenerative-classification](https://www.kaggle.com/competitions/rsna-2024-lumbar-spine-degenerative-classification) and https://mira.rsna.org/dataset/6. MRIs are provided in Digital Imaging and Communications in Medicine (DICOM) format along with comma-separated values (CSV) files which include the stenosis annotations and series descriptions for each patient (Appendix A Data Structure).

Discussion

We curated and created an expert annotated large MRI lumbar spine database, which represents the largest publicly available dataset for lumbar spondylosis. This rich dataset has further potential utility for future investigators including evaluation of intervertebral disc and vertebral endplate degenerative changes, which were beyond the scope of our competition.

We chose to have relatively loose inclusion criteria with respect to MRI protocols allowing for multiple "T2-like" sagittal sequences including conventional T2, STIR, and Dixon sequences and variable axial T2 acquisitions. This was done strategically in an effort to increase the generalizability of models trained on diverse imaging protocols.

Our dataset is not only unique in size but also in the diversity of data offered by the large number of contributing institutions and number of expert annotators. Also, few datasets include assessment of subarticular stenosis which has long been shown to be a cause of back pain and often associated with failed lumbar spine surgery.[8] Of the publicly available datasets, the Genodisc dataset most closely matches the current dataset in the depth of assessment and number of patients.[9] The Genodisc dataset includes only sagittal images but includes 2,287 studies across multiple imaging centers across Europe. One of the limitations of the Genodisc dataset as well as many other lumbar spondylosis datasets is low representation of high-grade disease by spinal level. We found that there was a very low natural incidence of moderate and especially severe stenosis at the L1/L2 and L2/L3 levels, which may limit model performance at these levels if the dataset is not supplemented for these minority classes.

One of the limitations in degenerative spine reporting is the inherent subjective nature of grading degenerative disease in the lumbar spine. We attempted to mitigate against subjective variation by incorporating a training manual for annotators and pre-annotation test dataset that utilized specific criteria based on published grading methodologies for spinal canal, neural foraminal, and subarticular stenosis (Appendix C). Also, we hoped having multiple annotators for each patient in the test sets would help increase consistency.

In summary, the RSNA Lumbar Spine Degenerative Classification dataset represents the largest, most geographically diverse, publicly available expert-annotated dataset of its kind. We hope this dataset will facilitate ML research development and improve outcomes in patients suffering from lumbar spondylosis. This dataset is made freely available to all researchers for non-commercial use.

Acknowledgements:

The authors would like to thank and acknowledge the contributions of Christopher Carr, MA, Sohier Dane, Maggie Demkin, MBA, and Michelle Riopel.

| Site | Sex Male | Female | Age (y) | Total Cases | At Least Moderate Narrowing | | | | Severe Narrowing | | | |
|---|---|---|---|---|---|---|---|---|---|---|---|---|
| | | | | | Neural Foramen | Subarticular | Spinal Canal | Any location | Neural Foramen | Subarticular | Spinal Canal | Any location |
| 1 | 99 | 135 | 55.8 +/- 10.9 (22-82) | 234 | 197 (84.1%) | 224 (95.7%) | 137 (58.5%) | 232 (99.1%) | 61 (26.0%) | 157 (67.0%) | 66 (28.2%) | 170 (72.6%) |
| 2 | 432 | 199 | 64.3 +/- 15.7 (18-89) | 631 | 520 (82.4%) | 564 (89.3%) | 349 (55.3%) | 594 (94.1%) | 244 (38.6%) | 368 (58.3%) | 206 (32.6%) | 435 (68.9%) |
| 3 | 110 | 163 | 48.4 +/- 14.7 (20-89) | 273 | 144 (52.7%) | 153 (56.0%) | 27 (9.89%) | 193 (70.6%) | 29 (10.6%) | 51 (18.6%) | 9 (3.29%) | 71 (26.0%) |
| 4 | 219 | 173 | 62.6 +/- 15.3 (19-89) | 392 | 321 (81.8%) | 348 (88.7%) | 230 (58.6%) | 368 (93.8%) | 163 (41.5%) | 221 (56.3%) | 131 (33.4%) | 264 (67.3%) |
| 5 | 160 | 235 | 56.6 +/- 14.1 (19-89) | 395 | 290 (73.4%) | 320 (81.0%) | 167 (42.2%) | 352 (89.1%) | 124 (31.3%) | 211 (53.4%) | 95 (24.0%) | 240 (60.7%) |
| 6 | 188 | 208 | 47.7 +/- 15.1 (18-85) | 396 | 190 (47.9%) | 227 (57.3%) | 40 (10.1%) | 283 (71.4%) | 35 (8.83%) | 69 (17.4%) | 11 (2.77%) | 89 (22.4%) |
| 7 | 42 | 52 | 59.7 +/- 14.0 (20-89) | 94 | 61 (64.8%) | 72 (76.5%) | 51 (54.2%) | 77 (81.9%) | 56 (59.5%) | 46 (48.9%) | 25 (26.5%) | 55 (58.5%) |
| 8 | 131 | 151 | 56.2 +/- 15.2 (18-86) | 282 | 199 (70.5%) | 228 (80.8%) | 111 (39.3%) | 251 (89.0%) | 75 (26.5%) | 123 (43.6%) | 61 (21.6%) | 153 (54.2%) |
| Total | 1,381 | 1,316 | 57.1 +/- 16.0 (18-89) | 2,697 | 1,922 (71.2%) | 2,136 (79.1%) | 1,112 (41.2%) | 2,350 (87.1%) | 761 (28.2%) | 1,246 (46.1%) | 604 (22.3%) | 1,477 (54.7%) |

Table 1. Patient demographics and prevalence of moderate and severe disease across the contributing institutions. At least moderate narrowing and severe narrowing columns refer to the presence of at least one spinal level of moderate or severe disease per imaging study.

|  |  | L1/L2 | L2/L3 | L3/L4 | L4/L5 | L5/S1 |
| --- | --- | --- | --- | --- | --- | --- |
| Neural Foramen | Moderate | 36 | 97 | 181 | 259 | 197 |
|  | Severe | 13 | 17 | 55 | 94 | 111 |
| Subarticular Recess | Moderate | 91 | 150 | 240 | 275 | 167 |
|  | Severe | 26 | 86 | 174 | 275 | 112 |
| Spinal Canal | Moderate | 35 | 66 | 89 | 73 | 22 |
|  | Severe | 8 | 30 | 65 | 79 | 20 |

Table 2. Patient demographics and prevalence of moderate and severe disease across the training, public test, and private test sets. At least moderate narrowing and severe narrowing columns refer to the presence of at least one spinal level of moderate or severe disease per imaging study.

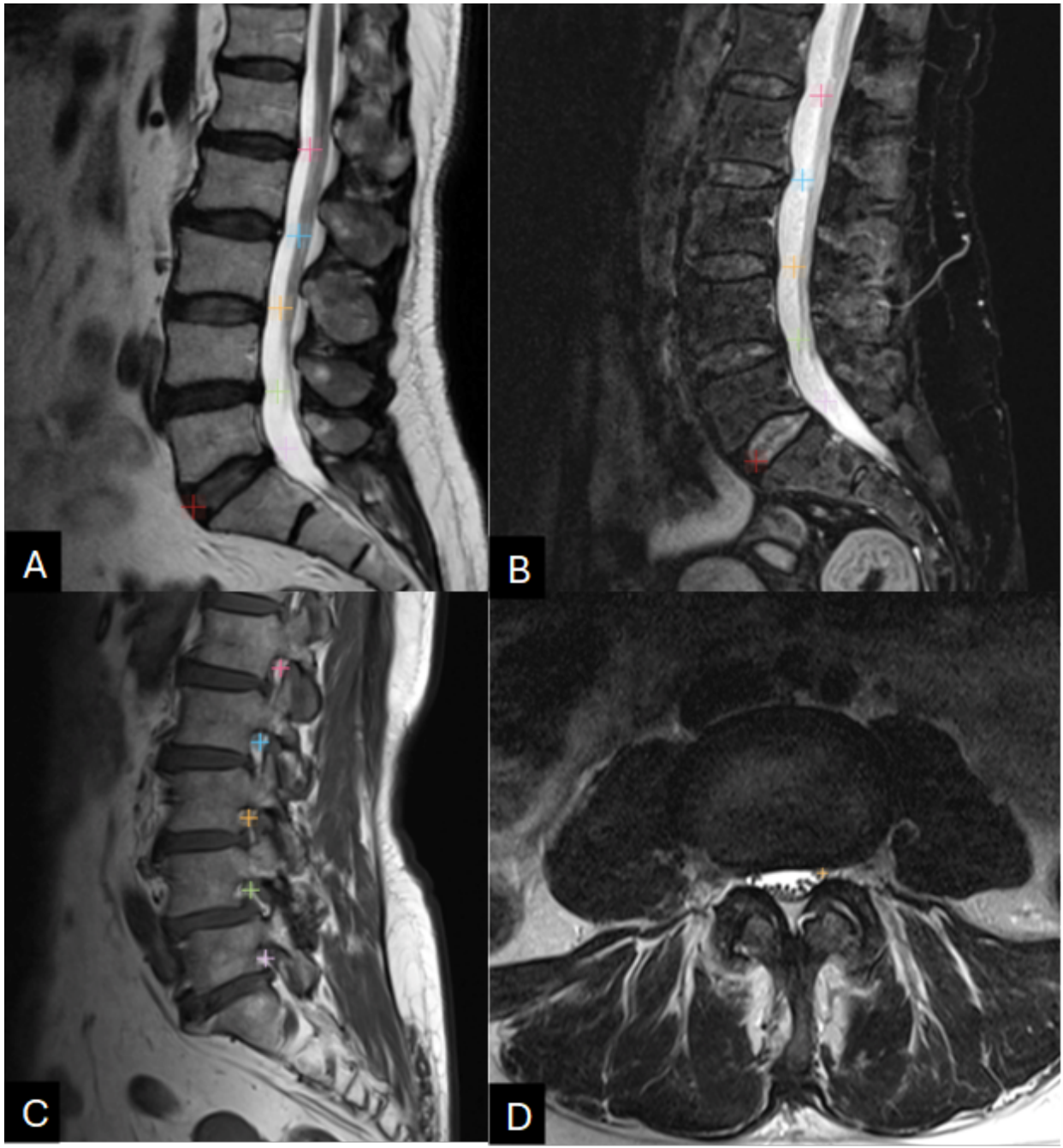

Figure 1. Images A (sagittal T2) and B (sagittal STIR) demonstrate the location of the localizers within the middle of the thecal sac at the level of the L1/L2 through L5/S1 intervertebral discs. These images also include the red localizer to demarcate the L5/S1 intervertebral disc level at the anterior margin of the disc. Images C (sagittal T1) and D (axial T2) demonstrate localizers centered within the left neural foramina and bilateral subarticular zones, respectively.

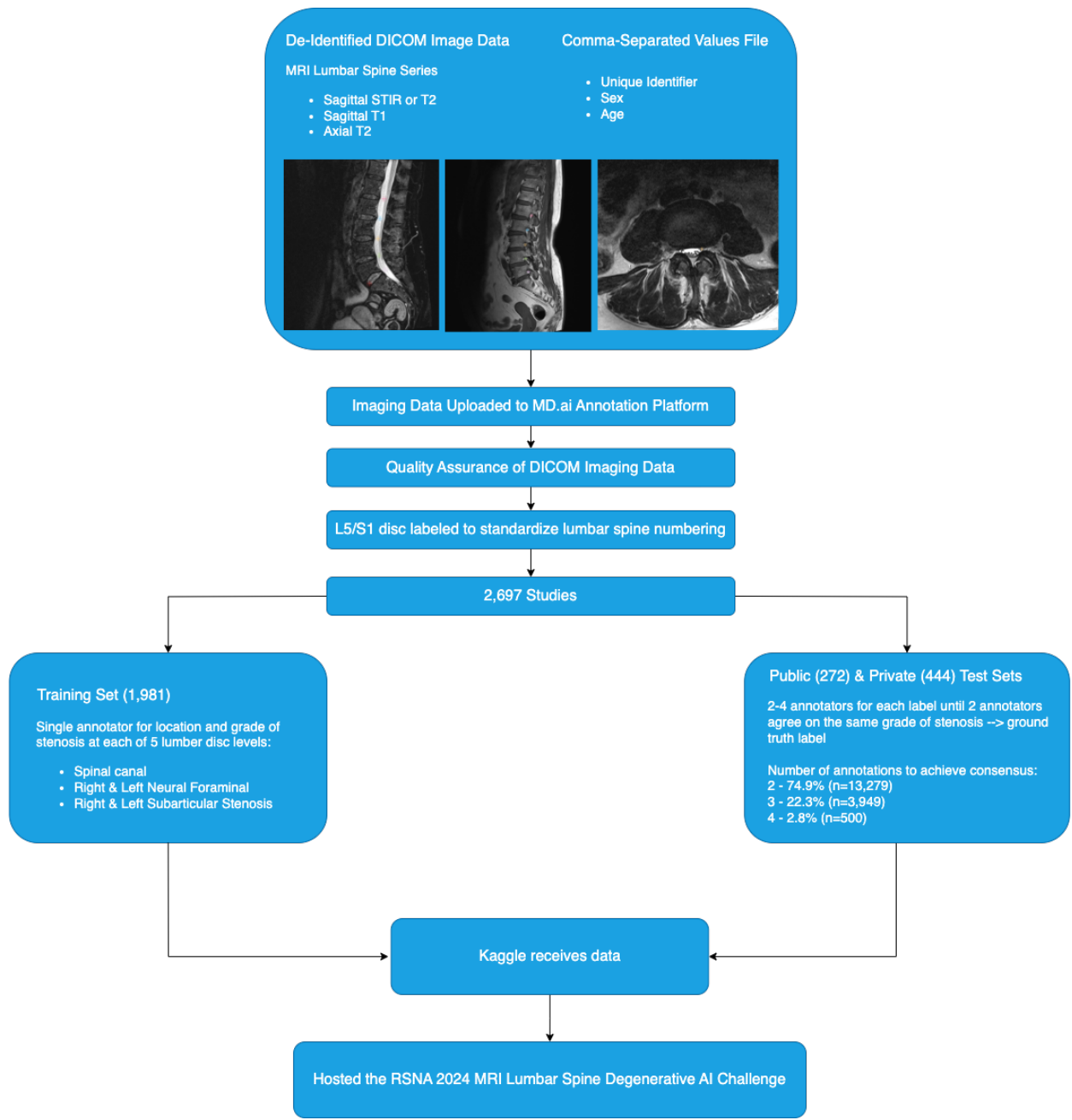

Figure 2. Flowchart summary of the data acquisition, curation, and annotation process.

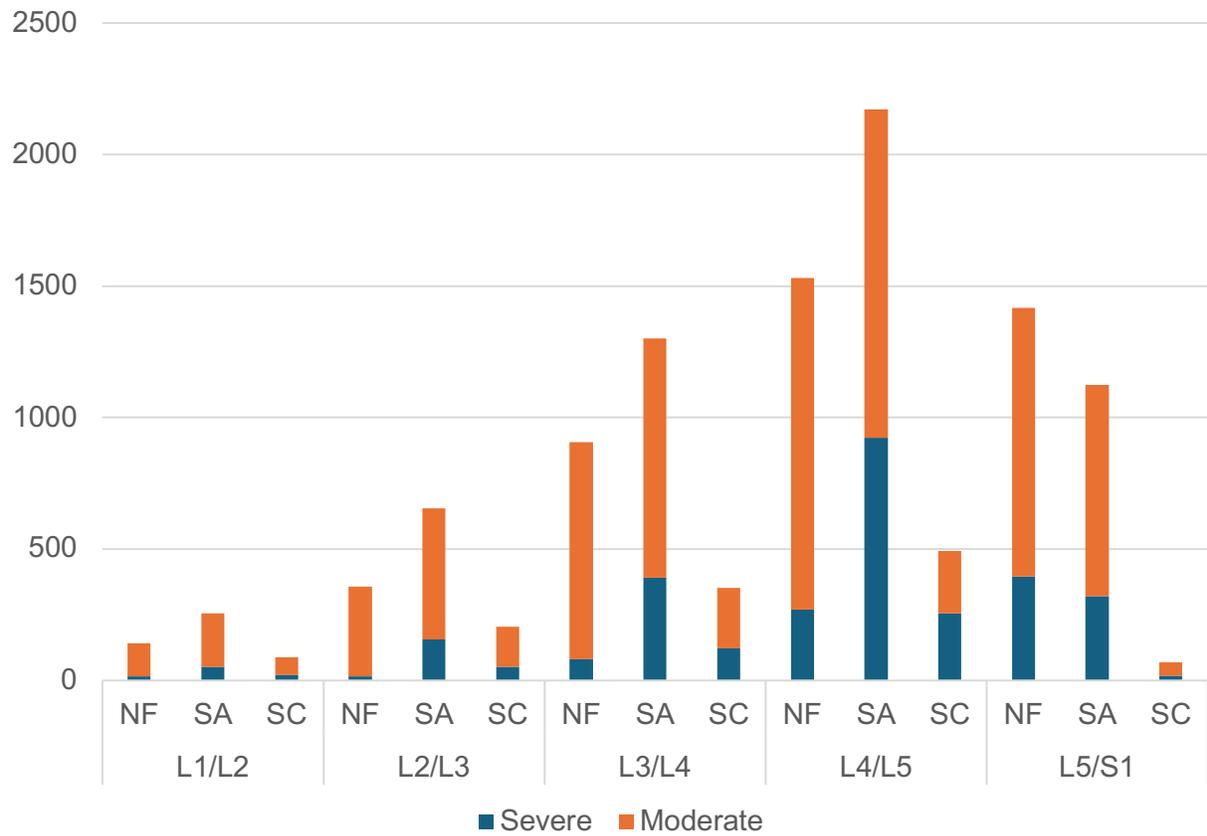

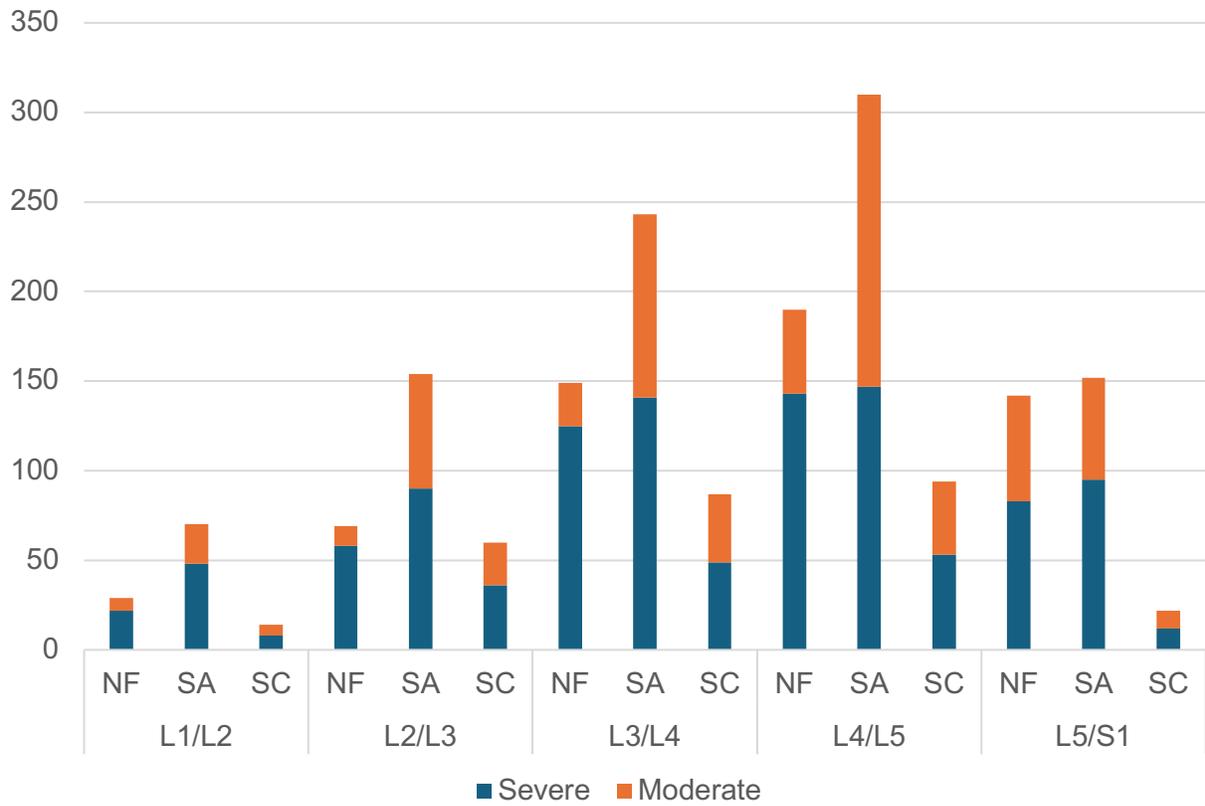

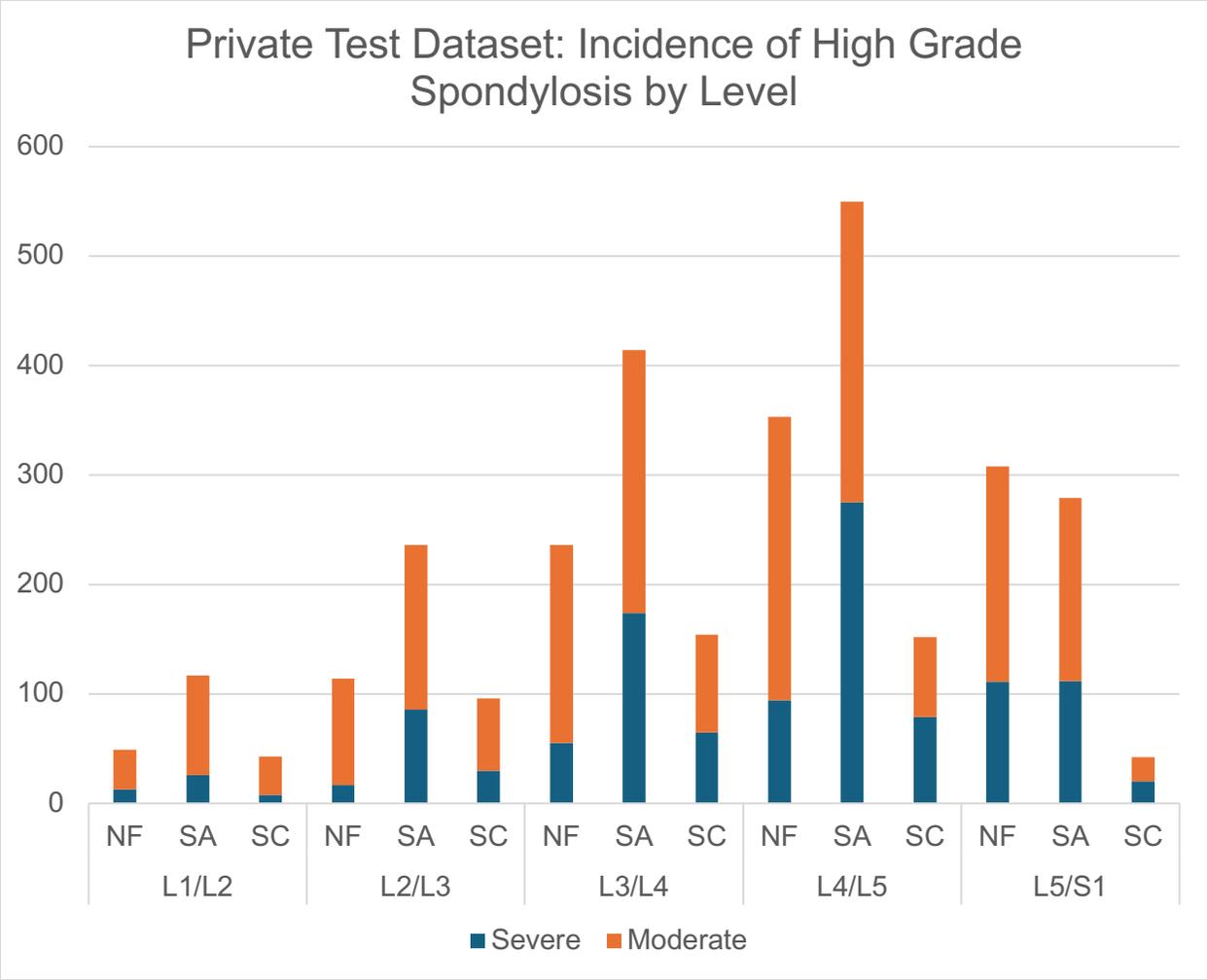

Figure 3. Distribution of moderate and severe disease across the training, public test, and private test sets. The y-axis represents the number of moderate and severe grades by level on either the right or left for neural foramina and subarticular recesses. NF = Neural Foramen; SR = Subarticular Recess; SC = Spinal Canal.

**Appendix A**

**Data Collection and Annotation**

The minimum acceptable dataset from each contributing site included sagittal T1, sagittal T2 fat saturated, and continuous axial T2 images through the entire lumbar spine. To manage substantial imaging protocol site variation with respect to sagittal T2 (no fat suppression at some sites) or angulation on axial T2 weighted images, we chose to include the variations to represent realistic acquisition differences in our dataset. A more diverse dataset theoretically should produce more generalizable models. We loosened criteria for a T2 sagittal fat suppressed sequence to accept a "T2-like" sagittal sequence which could include sagittal STIR, T2 fat-saturated, T2 spin echo, or Dixon sequence and included any axial T2 weighted sequence if it covered at least 3 intervertebral disc levels. We asked the contributing sites to provide both sagittal STIR and T2-weighted series as well as continuous and intervertebral disc level axial T2 weighted series if they had both to increase the utility of the dataset for future research. Since the goal of the challenge was to evaluate lumbar spondylosis exclusively, we asked that the contributing sites limit their studies to outpatient non-contrast studies which we hoped would filter out many patients with acute pathology including trauma, infection, and neoplastic processes. We excluded patients with postoperative hardware in the lumbar spine to avoid the limitations of susceptibility artifact from the hardware on evaluation especially in the neural foramina and subarticular recesses. If the patient had prior lumbar spine surgery without hardware, they could be included in the dataset. To standardize patient identification, we asked sites to collect the patients retrospectively starting with their most recent outpatient MRI lumbar spine studies and working backwards until they reached their contribution cohort of 400 patients. No patient was to be represented more than once in the submitted studies.

To facilitate cohort building and the annotation process, each site provided "pre-labels" for each study that designated the worst level and grade of spinal canal stenosis and then the grade for spinal canal, neural foraminal, and subarticular stenosis at that level according to the radiology report. This was done to survey the distribution of degenerative disease from each site prior to beginning the annotation process. It became apparent early on that the incidence of high-grade lumbar spondylosis was considerably lower at L1/L2, L2/L3, and to a lesser extent at L5/S1 compared to L3/L4 and L4/L5. To boost representation of these minority classes, 3 of the contributing sites supplemented the dataset with examples of high-grade disease at these under-represented levels. We asked contributing sites to include basic demographic information including patient age and sex. Contributing sites uploaded their data set as de-identified MRI images in DICOM format. Details about compliance of local ethics regulations, scan identification, and de-identification were left to the discretion of each contributing site and are described in Appendix B.

**Technical and Visual Quality Control before Annotation**

Prior to the volunteer annotators starting their annotation process, the L5/S1 intervertebral disc was labelled by a member of the organizing committee. This was done by placing a localizer labeled "L5/S1 disc" over the anterior aspect of the L5/S1 intervertebral disc at midline on the sagittal STIR, or if not present, whichever "T2-like" sagittal sequence that was included in the dataset. The most caudal fully formed disc space was labelled as L5/S1. A fully formed disc space is defined by a disc completely separating 2 adjacent vertebral bodies with at least partially formed circumferential annulus fibrosis and central nucleus pulposus. This label with localizer was included for both the training and test sets to limit the impact error in numbering the lumbar spine by both the annotators and trained models. During the initial labelling of the L5/S1 intervertebral disc level, each scan was visually inspected by a member from the organizing committee to flag studies that did not meet the requirements to be included in the dataset. This included flagging any studies for incomplete field of view on the sagittal images (not covering the majority of the bilateral neural foramina), substantial motion degradation, incomplete set of series (i.e. did not have the minimum of sagittal "T2-like" and T1 and axial T2 sequence requirements), patients with surgical hardware, or patients with a large mass within the spinal canal (e.g. tumor, abscess, or hematoma).

**Annotation**

Recruitment for expert annotators was accomplished through mass emails by the Radiological Society of North America (RSNA) to the membership of the American Society of Neuroradiology (ASNR) and American Society of Spine Radiology (ASSR). Annotators were required to be either neuroradiologists or musculoskeletal radiologists that frequently read outpatient lumbar spine MRIs. Second year neuroradiology fellows that had completed an ACGME-accredited 1 year neuroradiology fellowship also qualified as annotators. A web-based annotation platform (MD.ai, New York, NY) which shares many features of a Picture Archiving and Communication System (PACS) workstation was used for both uploading the imaging from each contributing site and used for annotation.

Prior to annotating for the AI challenge, each volunteer annotator was given an instruction manual detailing the grading schema (Appendix C) and a short instructional video on how to perform annotations on the annotation platform. Next, each annotator completed a set of 10 training cases in which they labelled spinal canal, right and left neural foraminal, and right and left subarticular recess stenosis at all 5 levels (L1/2 through L5/S1) for a total of 25 annotations per case. Each annotator's labels on the training cases were compared to consensus grades established by the neuroradiologists on the organizing committee. After achieving a minimum passing score (60% agreement with committee consensus grades) for at least one of the spinal canal, neural foraminal, or subarticular stenosis classes, the volunteer was assigned to one of the annotation tasks in which they achieved a passing score. Limiting each annotator to one annotation task at a time was found to be more efficient and less error prone than having each annotator switch between tasks. Rather than assigning a fixed number of cases to annotators, annotations were performed in a "crowdsourcing" manner where volunteers were free to determine how many studies they wanted to annotate. A public scoreboard that ranked annotators was used to motivate the volunteers.

For each study, the severity of spinal canal stenosis was labeled on the sagittal "T2-like" sequence that had the L5/S1 disc localizer, the neural foraminal stenosis was labeled on the sagittal T1 sequence, and the subarticular stenosis was labelled on the axial T2 series. The annotators were instructed to annotate using a localizer in the spinal canal at middle in the thecal sac at the level of the intervertebral disc, the neural foramen within the central aspect of the neural foramen, and the subarticular recess within the subarticular recess on the axial image that the annotator was using to make their severity grade (Figure 1).

To boost agreement, the four-point scale was contracted to three by combining the normal and mild grades to form a three-point scale of normal/mild, moderate, and severe. Next, the studies were separated into training (n=1,981), public test (n=272), and private test (n=444) sets. Test set cases were selected using stratified random sampling stratified on contributing site, age group, sex, and severity at each lumbar spine level. The cases with moderate and severe stenosis at L1/2, L2/3, and L5/S1 were augmented in the 2 test sets compared to the training set to ensure that there were enough representative examples in the public and private test sets to ensure that the competitors models would be tested throughout the range of severities at all levels including those levels of naturally lower incidence of severe stenosis. Data for the training set was annotated only once.  Each study in the public and private test set was annotated an additional 1-3 times until there was agreement between 2 annotators on the stenosis grade for each annotation, which established the final grade for each annotation.

**Technical and Visual Quality Control After Annotation**

After the annotation process was complete, custom Python scripts were written to evaluate for any potential annotation errors for the annotations for each case. These scripts checked for right and left mismatches between the localizer and the annotation label, spatial outliers for the annotation localizers, or missing annotations. Scripts were also written to create a collage of all the images for each sequence for each study to evaluate for any potential exclusion criteria, including missed surgical hardware and missing, flipped, or out or order images and to evaluate for incorrect imaging sequence labels. Each case that was flagged by the scripts as a potential annotation error was manually evaluated and corrected if possible or excluded from the study.

**Data structure**

MRI images in DICOM format are organized with the following directory hierarchy: [patient_id]/[series_id]/[image_instance_number].dcm.

Study level injury annotations are provided in CSV files. The study_id is a unique study level identifier for each study. Within each study, there are also unique series_id for each of the 3 series included for each study. The instance_number corresponds to the Instance Number metadata element and indicates the position of an image within the series. The train_series_descriptions.csv contains information regarding the imaging plane and sequence for each series_id. These include "Sagittal T2/STIR", "Sagittal T1", and "Axial T2". The train.csv file contain information about the grade ("Normal/Mild", "Moderate", or "Severe") with a column for spinal canal, right and left neural foraminal, and subarticular stenosis for each level of the spine including L1/L2, L2/L3, L3/L4, L4/L5, and L5/S1 for a total of 25 grading columns for each study_id. For example, the grade for left neural foraminal narrowing at L1/L2 is located in the column "left_neural_foraminal_narrowing_l1_l2". The train_label_coordinates.csv file has data including the location of the localizer corresponding to each stenosis grade provided by the annotators. Annotators were instructed to place a localizer for each stenosis grade centered within the spinal canal, neural foramen, or subarticular recess at the level of the intervertebral disc. This CSV file includes columns including the study_id, series_id, instance_number, condition, level, and [x/y] which gives the x/y coordinates for the center of the area that defined the label. The instance_number corresponds with the number corresponding to the DICOM file in the respective series (i.e. 1.dcm would be "1" in the csv file). The condition refers to spinal canal stenosis, right and left neural foraminal narrowing, and right and left subarticular stenosis. Level corresponds to the lumbar intervertebral disc level (e.g. "L1/L2"). Each row corresponds to each individual stenosis grade label. Examples of these tables can be found at

https://www.kaggle.com/competitions/rsna-2024-lumbar-spine-degenerative-classification/data.

**Data Processing for Kaggle Competition**

Due to the heterogeneity of the dataset with different metadata features at each contributing site, data harmonization is crucial for standardizing data input. A whitelist approach was used to address this step by restricting the number of DICOM elements and acts to perform another step of de-identification. The selected DICOM elements to retain were: Bits Allocated, Bits Stored, Columns, High Bit, Instance Number, ImageOrientationPatient, ImagePositionPatient, PatientPosition, PhotometricInterpretation, PixelRepresentation, PixelSpacing, RescaleIntercept, RescaleSlope, RescaleType, Rows, SamplesPerPixel, SliceLocation, SliceThickness, SpacingBetweenSlices, WindowCenter, WindowWidth. All identifying DICOM elements, including Patient ID, Study Instance UID, Series Instance UID, and SOP Instance UID were further de-identified using the pydicom Python library. The original image transfer compression method was kept, most if not all of which underwent compression using the Run-Length Encoding Lossless (RLELossless) method.

**Appendix B**

**Study Identification, Data Extraction, and Data De-Identification**

Contributing sites listed here do not correspond to the order in Table 1.

**Chiang Mai University, Thailand**

The Faculty of Medicine at Chiang Mai University searched their PACS backup archive (Synapse Radiology PACS version 5.7.000; FUJIFILM Medical systems) using RIS (Envision.Net) for MRIs of the lumbar spine without contrast. The search was restricted to outpatient status, patients 18 years of age or older, and the date range was between January 1, 2022 and August 15, 2023. Each study was manually reviewed by two radiologists to confirm that the inclusion criteria was met and cases were excluded according to the exclusion criteria. For each case, the worst level of spinal canal stenosis, right and left neural foramina narrowing and right and left subarticular narrowing were recorded. The patient age and sex were also recorded. The sagittal T1, sagittal T2, sagittal STIR or T2 with fat suppression, and axial T2 series for each study were exported from the PACS (IntelliSpace PACS; FUJIFILM Medical systems) to a local workstation within the hospital's network in DICOM format. The studies were anonymized and then uploaded to MD.ai for annotation.

**Clinical Center of University of Sarajevo, Bosnia and Herzegovina**

A comprehensive retrospective search was conducted using institutional Radiology Information System and PACS (IMPAX) by institutional procedural codes for MRI of the lumbar spine without contrast. The search included studies acquired from January 2020 until July 2023. Rigorous analysis of MRI reports was undertaken to confirm that the inclusion criteria was met and cases were excluded according to the exclusion criteria. The process of evaluation was performed by two experienced radiologists as well as the site's primary investigator. Demographic data were recorded for all cases, and for each case, the worst level of spinal canal stenosis, right and left neural foramina narrowing and right and left subarticular narrowing were recorded. Sagittal T1, sagittal T2, sagittal STIR, and axial T2 series for each study were exported from the PACS to a local workstation. All studies were anonymized and then uploaded to the platform.

**Gold Coast University Hospital, Gold Coast, Queensland, Australia**

Following approval by the Gold Coast Hospital and Health Service, a PACS search to identify all outpatient MRI lumbar spine studies up to June 30, 2023 was performed. The study reports were manually reviewed and studies removed if they met exclusion criteria, namely patient age < 18 years, presence of lumbosacral spinal hardware, active non-degenerative pathology, severe scoliosis, or presence of technical artifacts that limited the diagnostic evaluation of the study. The most recent studies were selected, to a total of 300. The level of worst spinal canal stenosis, right and left neural foraminal narrowing and right and left subarticular recess narrowing were recorded, in addition to patient age, gender and ethnicity where available. The studies were downloaded from the PACS in DICOM format and underwent de-identification prior to uploading into the RSNA database.

**Koc University School of Medicine, Istanbul, Turkey**

With the approval of the Koç University institutional review board, a retrospective search was performed using the institutional PACS (Sectra IDS7; Sectra AB) for MRI of the lumbar spine without contrast obtained between November 2022 and July 2023. Each study was manually evaluated to ensure that it met the inclusion criteria. For each case, the worst level of spinal canal stenosis, right and left neural foraminal narrowing and right and left subarticular narrowing were recorded. In all cases, data including patient age, patient sex and ethnicity (if available) were also collected. The sagittal T1, sagittal T2, sagittal STIR, and axial T2 series were anonymized and manually exported from the PACS as DICOM images. The images were then uploaded to MD.ai for final curation and processing.

**Thomas Jefferson University, USA**

Following approval of the Thomas Jefferson University institutional review board, a corpus of radiology reports between 2019 and 2023 was extracted from the dictation system (Nuance Powerscribe, Burlington, Vermont) using the mPower natural language processing search tool. The searches were filtered for outpatient encounters only using a single examination code (MRI LUMBAR SPINE WO CONTRAST) over a four year period for 15 hospitals in the Jefferson network. The impression section of the reports were specifically searched for a designated level (e.g. L1-2) and the word "stenosis" and the absence of the words instrumentation, fusion, construct, screws, rods, ACDF, cage, spacer, laminectomy, or decompression. A total of 1000 examples at each level were identified. These were randomized and the top 100 from each lumbar level were aggregated into one list yielding 500 studies. Studies were queried and downloaded from the Hyland VNA and de-identified using CTP Anonymizer and uploaded to the RSNA MD.ai endpoint.

**Diagnósticos da América S.A. (DASA), São Paulo, Brazil**

We queried our PACS system for studies with the modality 'MR,' the study description 'RM COLUNA LOMBAR,' and a date range from January 8, 2023, to March 22, 2023. The cases were manually reviewed based on the inclusion criteria. Afterward, the selected cases were anonymized using RSNA CTP software before being shared.

**University of California San Francisco, USA**

The primary site investigator searched their institutional RIS (Radiant; Epic Systems Corporation) using Nuance mPower (Nuance Communications) by the current procedural terminology (CPT) code 72148 which codes for MRI of the lumbar spine without contrast. The search was restricted to outpatient status, patients 18 years of age or older, and the date range was restricted to a two month date range (09/01/2023 to 10/31/2023). Each study was manually reviewed to confirm that the inclusion criteria was met and cases were excluded according to the exclusion criteria. Cases meeting inclusion criteria were securely downloaded using a self-service automated image retrieval platform (UCSF AIR) utilizing automated de-identification of header data and stored in a secure server within the hospital's network. For each de-identified case, the worst level of spinal canal stenosis, right and left neural foraminal narrowing and right and left subarticular narrowing were recorded. The patient age, sex, and ethnicity (if available) were also recorded. Anonymized relevant sequences (sagittal T1, sagittal T2 fat suppressed, and axial T2) were then securely uploaded to MD.ai for annotation.

**University of Utah, USA**

The primary site investigator searched their institutional RIS (Radiant; Epic Systems Corporation) using Nuance mPower (Nuance Communications) by the current procedural terminology (CPT) code 72148 which codes for MRI of the lumbar spine without contrast. The search was restricted to outpatient status, patients 18 years of age or older, and the date range was restricted to within one month of the database query (03/22/2023 to 04/21/2023). Each study was manually reviewed to confirm that the inclusion criteria was met and cases were excluded according to the exclusion criteria. For each case, the worst level of spinal canal stenosis, right and left neural foraminal narrowing and right and left subarticular narrowing were recorded. The patient age, sex, and ethnicity (if available) were also recorded. The sagittal T1, sagittal T2, sagittal STIR, and axial T2 series for each study were exported from the PACS (IntelliSpace PACS; Philips Healthcare) to a local workstation within the hospital's network. The studies were anonymized and then uploaded to MD.ai for annotation.

**Appendix C**